\documentclass[letterpaper,10pt]{article}
\usepackage[letterpaper,left=1.5in,right=1.5in,top=1in,bottom=1in]{geometry}
\usepackage{stix}
\usepackage{setspace}
\usepackage[small]{titlesec}
\usepackage[utf8]{inputenc} 

\usepackage[american]{babel} 
\usepackage{csquotes} 
\usepackage[babel=true]{microtype} 
\usepackage{xspace}

\usepackage{graphicx}

\usepackage{caption}
\addto\captionsamerican{} 

\usepackage{booktabs}

\usepackage{mathtools}
\usepackage{siunitx}
\usepackage{stix}

\usepackage{cite}

\newcommand{\Hair}{\ifmmode\mskip1mu\else\kern0.08em\fi}

\usepackage{hyperref}

\usepackage{csvsimple}

\begin{document}

\title{Fundamental Challenges of Cyber-Physical Systems\\ Security Modeling}

\author{Georgios~Bakirtzis \and 
  Garrett~L.~Ward \and
  Christopher J. Deloglos \and
  Carl~R.~Elks \and
  Barry M. Horowitz \and
  Cody~H.~Fleming%
\thanks{G. Bakirtzis, B.M. Horowitz, and C.H. Fleming are
    with the University of Virginia, Charlottesville, VA USA.
     E-mail: \{bakirtzis,bh8e,fleming\}@virginia.edu}
\thanks{G.L. Ward is with Honeywell International Inc.
    Email: garrett.ward@honeywell.com}
\thanks{C.J. Deloglos, and C.R. Elks are 
with Virginia Commonwealth University, Richmond, VA USA.
    Email: \{delogloscj,crelks\}@vcu.edu}
}
\renewcommand\footnotemark{}
\date{}

\maketitle
\thispagestyle{plain} 
\pagestyle{plain}

\begin{abstract}
  Systems modeling practice lacks security analysis tools
  that can interface with modeling languages
  to facilitate \emph{security by design}.
  Security by design is a necessity 
  in the age of safety critical cyber-physical systems,
  where security violations can cause hazards.
  Currently, the overlap between security
  and safety is narrow.
  But deploying cyber-physical systems
  means that today's adversaries can intentionally trigger accidents.
  By implementing security assessment tools
  for modeling languages we are better able
  to address threats earlier in the system's lifecycle
  and, therefore, assure their safe and secure behavior
  in their eventual deployment.
  We posit that cyber-physical systems security modeling
  is practiced insufficiently
  because it is still addressed similarly to information technology systems.
\end{abstract}

\section{Introduction}

Cybersecurity has emerged as an important property
that can affect the reliable, dependable,
and safe operation (to name but a few ``-ilities'' security affects) of cyber-physical systems (CPS)~\cite{nicol:2004}.
This is primarily because of the potential disruptions in service
that an attack vector imposes on such systems,
which operate in the physical environment and,
therefore, can be transitioned to a hazardous state externally.
To manage this risk, we must start assessing security threats
as early as possible in a system's lifecycle and then throughout
with the use of models.
Such models should be augmented to be a living artifact of security considerations in addition to design choices.

Modeling has played a vital role in assuring the expected operation
of systems in the areas of reliability, dependability,
and safety assessment.
The benefits of modeling in the design process have allowed
a \emph{what if} analysis of component failures, dependable modes,
and hazardous conditions in systems that are not yet implemented, thereby
saving time, reducing costs, and providing guarantees, whilst managing risk.

Additionally, reliability, dependability, and safety analysis have been assisted
by integrating ``-ility'' specific assessment tools within systems engineering modeling tools.
By doing so, systems engineers gain access and insight into those disciplines
without necessarily being experts in each individual area.
Tools, then, are the catalyst to incorporating different types
of analysis in the model-based engineering view of system design.
For CPS, it is important to integrate security analysis tools
with languages that can model the interaction between software-controlled behavior and physical consequences.

This integration is key in CPS, where digital systems control physical behavior
and therefore are tightly coupled.
However, cybersecurity tools often omit such interactions 
and consider only mitigation strategies related 
to hardening solutions and data flow but not physical value integrity.
This defines an important gap between system modeling,
security, and ultimately safety: security tools are disjoint from behavioral consequences.
For example, modeling attacks in Microsoft's threat modeling tool 
or attack trees assumes that the system must be a collection of IT infrastructure
with no physical interactions.
This assumption does not hold for CPS, where undesired physical consequences are the primary loss
we mitigate against regardless of the nature of its origin (intrinsic safety fault or attack).
This general omission in threat modeling tools can only be attributed to viewing CPS as IT systems, a view that does not recognize the new challenges CPS pose by interacting with the environment.

Although the theory and visualization techniques
for transitioning to model-based cybersecurity analysis are advancing~\cite{noel:2016,jauhar:2015,dwivedi:2018,huff:2019,bakirtzis:2019,bakirtzis:2018},
there are still several challenges.
The first is moving to mature software implementation
of security modeling tools that can be run
within already existing systems modeling tools.
The second relates to a more general goal, but perhaps more important,
of transitioning security into the systems engineering process such that
threats and vulnerabilities are addressed in collaboration
between systems engineers, security analysts, and safety experts.
This transition would allow both a common language between systems designers
and security engineers but also for systems engineers
to be aware of possible cybersecurity violations
without necessarily being security analysts themselves.
In turn, this language and awareness would increase the efficiency and effectiveness
of building security by design
and allow modeling and relating both attacks and their physical consequences.

We show a possible research trajectory
for addressing these challenges
by implementing prototype tools
that interface CPS modeling
and security.
In the coming years more vendors
will build isolated tools to address security
within a systems modeling paradigm,
which makes this research program timely and necessary.

\section{Transitioning to Model-Based Cybersecurity for Cyber-Physical Systems}


We posit that a collection of capabilities is necessary
to promote a systematic and prescriptive approach to security.
As a \emph{first step} to integrating security analysis tools
in the systems modeling process we require three capabilities:
\begin{enumerate}
    \item export modeling language-specific systems models to a general architectural model;
    \item associate attack vector data to the general model; and
    \item a visual display of both system models and attack vectors in a common graphical user interface to enable analysis and decision making.
\end{enumerate}

The intellectual framework necessary for such capabilities is not straightforward.
Decisions have to be made about how to allow modularity
between modeling languages, what system model fidelity is necessary for security analysis
to be useful in the early lifecycle, and how to manage
and filter the large amount of attack vector data produced
at early stage.
The results of this research can guide future attempts
in designing and developing security analysis tools
for any given systems modeling tool
but they also nudge towards \emph{how} we should be modeling CPS
for assessing cyber-physical security early in the lifecycle
before implementation and deployment.
The second result is also applicable 
to the documentation and reevaluation
of already deployed CPS in terms of their security posture.

As the complexity of system models increases
and the need for security analysis becomes commonplace
such a toolkit will be indispensable to the system designer.
The rise of security violations in CPS indicates
that the sooner we familiarize system designers
with security workflows, the closer we will be
to addressing security early in the systems lifecycle
and throughout in its later stages through the use of models.

The main problem this work addresses is
how to design tools with associated ``blue team'' practices
that find relevant attack vectors
on a system when the only available description is design documents
or incomplete documentation of legacy systems.
This is particularly difficult because a large number
of security tools used in the industry are assuming a realized system.
For example, it is common to use static analysis tools which run
on source code and by definition can only be used after implementation.
Likewise, dynamic analysis tools require a working device which is not available in the design phase.
Other threat modeling tools, such as Microsoft threat modeling tool, can be used
at the design phase, but they are primarily focused on the IT infrastructure and, therefore, are insufficient for assessing security in CPS.

This narrow focus does not allow for the modeling of the physical interactions
with the system under design and, therefore, cannot map threats
to environmental consequences.
Tools based on attack trees are often used to augment results
from such threat modeling.
Therefore, they are also focused on the risk to the IT infrastructure
and not the risk of causing undesirable physical behaviors.
We posit that this view of security is insufficient for CPS.

Instead, transitioning to model-based security assessment
within a systems modeling tool expands the notion
of security by giving access
to system assets, their relationships, and the physical behaviors.
Having access to comprehensive system models allows any security analysis
to relate attacks to components, interactions, and potential operational losses.

In recent years, security modeling practice has moved
from a perspective of hardening a list of assets
to representing things as graphs, 
which is congruent with how attackers operate
in reality~\cite{lambert:2015}.
However, security modeling is often disjoint
from system modeling lacking connection to both the system's physical consequences and also to its requirements.

To overcome this challenge, we posit that the system model must include extra design information at the earliest possible stage in the system's lifecycle
in addition to current modeling practice~\cite{bakirtzis:2018b}.
This extra design information takes the form
of an initial architecture, which includes specific information
about the eventual design of the system.
The initial architecture need not be an exact match
to the eventual deployed system
but it must reflect some implementation of the desired system with the expectation that it will
changes through design iterations.

The inputs to the security tools are the system model
and security data in the form of natural text.
Particularly, the attack vectors consist of databases containing vulnerability, weakness, and attack pattern data, such as the ones published by MITRE.

High-level descriptions of system components
and interactions will tend to match attack pattern and weakness instances; low-level
or more specific descriptions of software
and hardware platforms will relate more closely to vulnerability instances.
Furthermore, each of these datasets contains interconnections
with one another which creates the possibility
of capturing both the attacker’s perspective from attack pattern
and the system owner’s perspective from weakness and vulnerability.
Without all these differing views of exploits our perspective
on the system’s security posture is incomplete.

The main output, then, is this association
of attack vectors to the system model.
By using the initial design of the system
and the attack vectors that associate with it
we can compute several commonly used metrics
in security analysis.

Security analysis is not a linearly iterative process
and at this stage it is common to apply architectural refinements
to the model.
What we mean by architecture refinement is the addition
of increasingly specific information
in the model such that the relevance
of attack vectors increases the closer we get to deployment.
This is important especially in the early lifecycle,
where requirements and design choices are not finalized
and the impact to cost is lowest and effectiveness highest.

We further posit that security analysis at this stage should be qualitative.
This is because quantitative information for cyber-physical attacks is limited
and ultimately nuanced expert input is necessary to understand how to interpret a metric to show that a system is more susceptible to attacks
over another.
For example, a common mistake is to use CVSS
as a potential metric for risk.
However, CVSS only defines severity of a given vulnerability and not risk \cite{CVSS}.
This is a general open problem in cybersecurity of not knowing what data we should record 
to enable us to use quantitative decision theory.
This implies that probabilistic approaches
to cybersecurity are limited because attacker behavior can be non-probabilistic
and we currently have limited data to learn the general patterns
of attacker behavior for such methods to be useful.




\section{Demonstration: Particle Separation Centrifuge}

\begin{figure*}[!th]
    \centering
    \includegraphics[width=1\linewidth]{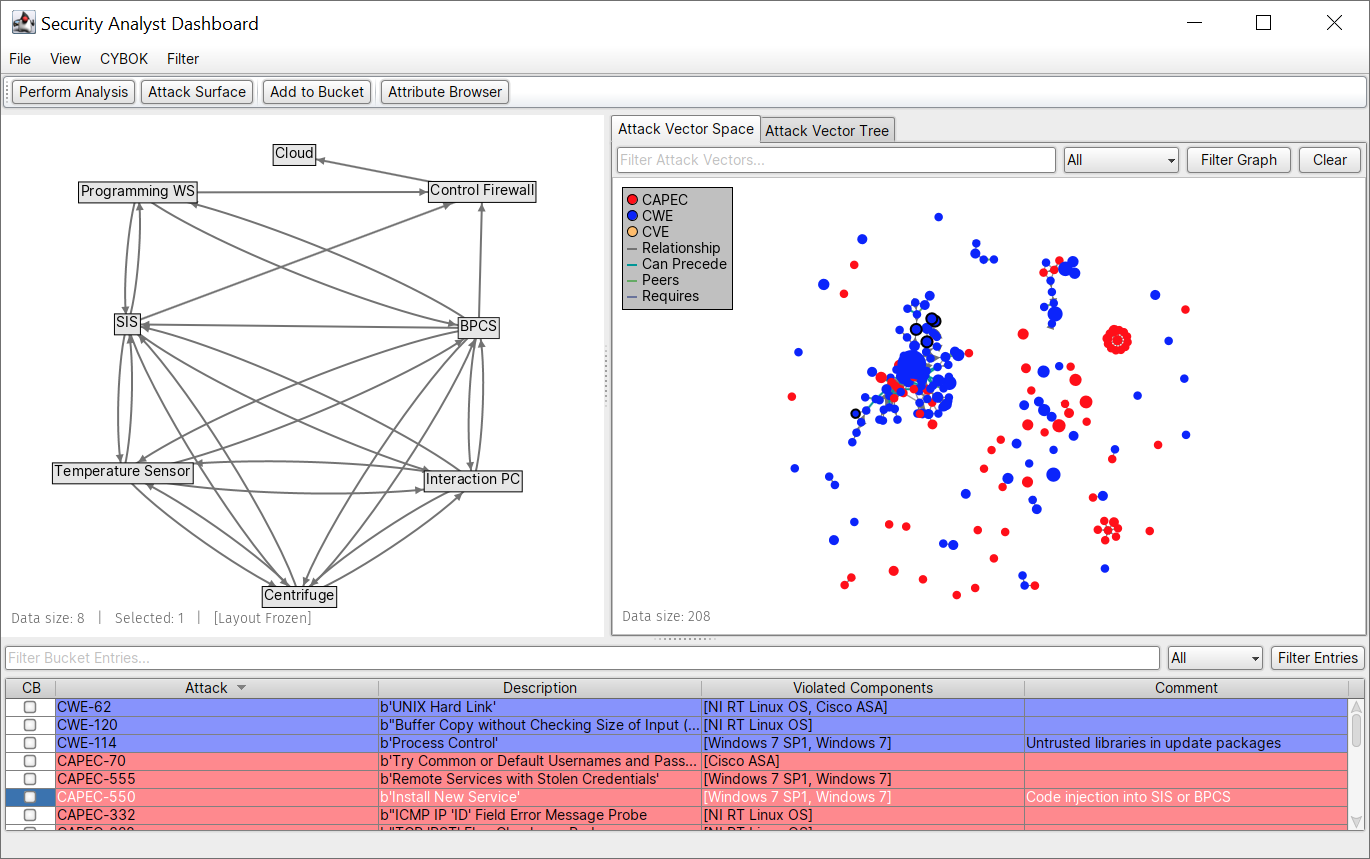}
    \caption{The demonstration merges system models with attack vector data to promote model-based security.}
    \label{fig:cs_whole_screen}
\end{figure*}

We have implemented prototype open-source tools to support this view
of security. These include an exporter from SysML to a graph system model \cite{graphml}, a search engine that takes as input a system model and outputs the associated attack vectors~\cite{cybok}, and a dashboard \cite{dashboard} that merges system modeling with the security data associated with it.

The use case we analyze is a real-world example of a particle separation centrifuge. Centrifugation is a technique used for the separation of particulate from a solution according to their size, density, and viscosity of the medium and rotor speed. In this application, the separation is highly sensitive to temperature. If the temperature is too low, the separation will not be productive and the result is a viscous product. If the temperature is too high, the chemical composition of the solution in the centrifuge tube can become unstable and cause an explosion/fire. If the rotor speed fluctuates beyond $\pm 20$ rpm of the set point the resultant product is not useful. Thus, the system is both functionally safety and security critical.

The supervisory control and data acquisition (SCADA) system components for the particle separation centrifuge are as follows (Fig.~\ref{fig:cs_whole_screen}).
\begin{itemize}
    \item \emph{Programming WS}: the controller of the centrifuge, programmed in NI LabVIEW, and monitored by operators. 
    \item \emph{Control firewall}: the firewall isolates the corporate network from the control network.
    \item \emph{SIS platform}: a redundant safety monitor for the centrifuge controller, for example, temperature is too high for commanded mode or speed is too high. 
     \item \emph{BPCS platform}: the main centrifuge controller interfaced through MODBUS.
    \item \emph{Temperature sensor}: a precision passive temperature probe that monitors the temperature of the solution to $\pm$ \SI{0.2}{\celsius}.
    \item \emph{Centrifuge}: a precision variable speed centrifuge capable of a maximal rotational speed of $10,000$ rpm and can regulate to within $\pm 1$ rpm of set point.
\end{itemize}

  Running the prototype tools shows that the total number of attack vectors returned by the search process is large (Table~\ref{table:cs_results}). Filtering functionality is implemented to manage these attack vectors~\cite{bakirtzis:2018}.
  But the general lessons stemming from the large result space is that it is highly sensitive to the fidelity of the model.
  If the model is closer to implementation; that is, it contains most information about hardware and software design choices the result space will be more specific.
  Another possible solution is to abstract away vulnerabilities at the earlier stages
  of the design lifecycle where the model is more abstract and therefore better relates to attack patterns and weaknesses.
  
    \begin{table}[!h]
	\caption{A fragment of attributes and the number of associated attack vectors from the SCADA model.}
	\begin{center}
		\begin{filecontents*}{Results.csv}
Attribute,Attack Patterns,Weaknesses,Vulnerabilities
Cisco ASA,2,1,3776
NI RT Linux OS,54,75,9673
Windows 7,41,73,6627
Labview,0,0,6
NI cRIO 9063,0,0,7
NI cRIO 9064,0,0,7
\end{filecontents*}
\csvautobooktabular{Results.csv}
	\end{center}
	\label{table:cs_results}
  \end{table}


  Different design information added to the model yield different result spaces depending on the generality of the model.
  In the dashboard we allow for the systems engineer or security analyst
  to change the model on the fly and immediately see the new results.
  The dashboard acts as a what-if analysis, where different architectures are evaluated by experts iteratively to lead to an acceptably secured system.
  The assertion here is that a component or subsystem that relates with less attack vectors
  than a functionally equivalent system has a better security posture.

Several relevant attack vectors were discovered for each of the subsystems.
For example, both the BPCS and SIS platforms were proposed 
of being vulnerable to CWE-78 -- OS Command Injection. This weakness describes an attack scenario where an upstream attacker may inject all or part of an operating system command onto an externally influenced input for the BPCS and SIS platforms disrupting or manipulating the platform's operation. This attack may result in compromised control of the centrifuge, manifesting in destruction of the manufactured product or damage to the centrifuge itself, which could cause accidents.
This is not an unreasonable scenario as is illustrated by Triton -- where malware was used to disable the safety systems of a petrochemical plant. Attack vectors can lead to unsafe control actions in CPS and must be addressed early on, but no \emph{science} of security exists yet to map attack vectors to physical consequences and leverage the existing power of systems modeling. 
 
An intrinsic limitation of this prototype approach is that it is grounded
in relating attack vectors to the system model through natural language processing.
This can lead to a very sensitive model in terms of how many attack vectors are produced depending on minor changes in attribute descriptions.
For example, system nodes with unspecific properties result in large numbers of attributes with many irrelevant results.
A more sophisticated modeling tool that enables and encourages systems engineers to add specific, security-related properties to the model without needing extensive domain-specific knowledge about security could mitigate this limitation.
Such a tool would permit the integration of security modeling throughout the development lifecycle of CPS.

\section{Conclusion}

We believe cybersecurity assessment capabilities should be part
of any modern systems modeling language.
Indeed, their full integration within a modeling language would be an important step
to promoting security by design.
Having security tools as part
of a modeling language would make them accessible
to systems designers.
In turn, this would increase the collaboration of security analysts
and systems engineers tackling the same problem within a common language.
In this position paper, we show that there are a number of challenges
to overcome to reach model-based security in practice.
The dependability community has a long tradition
of promoting a tool-based approach to its field,
which should be extended to security of CPS at the earliest stage
of the lifecycle possible.

Specifically, the assurance of a number of ``-ilities'', 
for example, reliability, dependability, and safety
is considered to be part the responsibility of the systems engineers.
However, security is often exercised as a separate problem,
to be tackled later in the lifecycle by security teams
and outside of the modeling domain.
We assert that this approach is insufficient to design and implement secure
and, therefore, reliable, safe, and dependable CPS.
The main result of this paper is the identification of the capabilities
which allow security analysis to become part of the systems engineering process
to precisely promote this view for CPS.
Through the demonstration of the capabilities needed to transition to model-based security analysis, we hope the tools industry will consider implementing accessible tools that enable systems designers who may not be security analysts to use these tools effectively in the design phase.
Of particular interest to industry is the evaluation
of any such tool with regards to its efficacy
to produce meaningful outputs.
This requires a number of resources unlikely
to be found in academia, for example, expert analysts,
diverse models, and ultimately confidence in the results.

\section{Acknowledgements}

This material is based, in part, upon work supported
  by the Stevens Institute of Technology through SERC 
  under USDOD contract HQ0034-19-D-0003. 
  SERC is a federally funded university affiliated research center managed 
  by Stevens Institute of Technology.
  Any opinions, findings and conclusions 
  or recommendations expressed in this material are those
  of the authors and do not necessarily reflect the views of USDOD.
\bibliographystyle{IEEEtran}
\bibliography{manuscript}

\begin{thebibliography}{10}
\providecommand{\url}[1]{#1}
\csname url@samestyle\endcsname
\providecommand{\newblock}{\relax}
\providecommand{\bibinfo}[2]{#2}
\providecommand{\BIBentrySTDinterwordspacing}{\spaceskip=0pt\relax}
\providecommand{\BIBentryALTinterwordstretchfactor}{4}
\providecommand{\BIBentryALTinterwordspacing}{\spaceskip=\fontdimen2\font plus
\BIBentryALTinterwordstretchfactor\fontdimen3\font minus
  \fontdimen4\font\relax}
\providecommand{\BIBforeignlanguage}[2]{{%
\expandafter\ifx\csname l@#1\endcsname\relax
\typeout{** WARNING: IEEEtran.bst: No hyphenation pattern has been}%
\typeout{** loaded for the language `#1'. Using the pattern for}%
\typeout{** the default language instead.}%
\else
\language=\csname l@#1\endcsname
\fi
#2}}
\providecommand{\BIBdecl}{\relax}
\BIBdecl

\bibitem{nicol:2004}
D.~M. Nicol, W.~H. Sanders, and K.~S. Trivedi, ``Model-based evaluation: From
  dependability to security,'' \emph{{IEEE} Transactions on Dependable and
  Secure Computing}, 2004.

\bibitem{noel:2016}
S.~Noel, E.~Harley, K.~H. Tam, M.~Limiero, and M.~Share, ``Cygraph: graph-based
  analytics and visualization for cybersecurity,'' in \emph{Handbook of
  Statistics}.\hskip 1em plus 0.5em minus 0.4em\relax Elsevier, 2016.

\bibitem{jauhar:2015}
S.~Jauhar, B.~Chen, W.~G. Temple, X.~Dong, Z.~Kalbarczyk, W.~H. Sanders, and
  D.~M. Nicol, ``Model-based cybersecurity assessment with {NESCOR} smart grid
  failure scenarios,'' in \emph{Proceedings of the 21st IEEE PRDC}.\hskip 1em
  plus 0.5em minus 0.4em\relax IEEE, 2015.

\bibitem{dwivedi:2018}
A.~Dwivedi, ``Implementing cyber resilient designs through graph analytics
  assisted model based systems engineering,'' in \emph{Proceedings of the 2018
  IEEE QRS-C}.\hskip 1em plus 0.5em minus 0.4em\relax IEEE, 2018.

\bibitem{huff:2019}
J.~Huff, H.~Medal, and K.~Griendling, ``A model-based systems engineering
  approach to critical infrastructure vulnerability assessment and decision
  analysis,'' \emph{Systems Engineering}, 2019.

\bibitem{bakirtzis:2019}
G.~Bakirtzis, B.~J. Simon, A.~G. Collins, C.~H. Fleming, and C.~R. Elks,
  ``Data-driven vulnerability exploration for design phase system analysis,''
  \emph{IEEE Systems Journal}, 2019.

\bibitem{bakirtzis:2018}
G.~Bakirtzis, B.~J. Simon, C.~H. Fleming, and C.~R. Elks, ``Looking for a black
  cat in a dark room: Security visualization for cyber-physical system design
  and analysis,'' in \emph{Proceedings of the 2018 IEEE VizSec}.\hskip 1em plus
  0.5em minus 0.4em\relax IEEE, 2018.

\bibitem{lambert:2015}
J.~Lambert, ``{Defenders think in lists. Attackers think in graphs. As long as
  this is true, attackers win.}'' \url{https://perma.cc/6NZ2-A2HY}, 2015.

\bibitem{bakirtzis:2018b}
G.~Bakirtzis, B.~T. Carter, C.~R. Elks, and C.~H. Fleming, ``A model-based
  approach to security analysis for cyber-physical systems,'' in
  \emph{Proceedings of the 2018 IEEE SysCon}.\hskip 1em plus 0.5em minus
  0.4em\relax IEEE, 2018.

\bibitem{CVSS}
J.~Spring, A.~H. E.~Hatleback, A.~Manion, and D.~Shic, ``Towards improving
  {CVSS},'' SEI, CMU, Tech. Rep., 2018.

\bibitem{graphml}
\BIBentryALTinterwordspacing
G.~Bakirtzis and B.~J. Simon, ``{GraphML} export,'' 2018. [Online]. Available:
  \url{https://doi.org/10.5281/zenodo.1308914}
\BIBentrySTDinterwordspacing

\bibitem{cybok}
\BIBentryALTinterwordspacing
G.~Bakirtzis, ``{CYBOK} command line interface,'' 2020. [Online]. Available:
  \url{https://doi.org/10.5281/zenodo.3766874}
\BIBentrySTDinterwordspacing

\bibitem{dashboard}
\BIBentryALTinterwordspacing
G.~Bakirtzis and B.~J. Simon, ``Security analyst dashboard,'' 2018. [Online].
  Available: \url{https://doi.org/10.5281/zenodo.1318537}
\BIBentrySTDinterwordspacing

\end{thebibliography}

\end{document}